# Experimental evidence of water formation on interstellar dust grains

F. Dulieu [1a], L. Amiaud [1], J-H. Fillion [1,b], E. Matar [1], A. Momeni [1], V. Pirronello [2], J. L. Lemaire [1]

[1] LERMA, UMR8112 du CNRS, de l'Observatoire de Paris et de l'Université de Cergy Pontoise, 5 mail Gay Lussac, 95031 Cergy Pontoise Cedex, France

[2] Università di Catania, DMFCI, Viale A. Doria 6, 95125 Catania, Italia

[a] Corresponding author : francois.dulieu@u-cergy.fr

[b] Present address: Université Pierre et Marie Curie - Paris 6, LPMAA, UMR 7092 du CNRS

**The synthesis of water is one necessary step in the origin and development of life. It is believed that pristine water is formed and grows on the surface of icy dust grains in dark interstellar clouds. Until now, there has been no experimental evidence whether this scenario is feasible or not. We present here the first experimental evidence of water synthesis under interstellar conditions. After D and O deposition on a water ice substrate ($H_2O$) held at 10 K, we observe production of HDO and $D_2O$. The water substrate itself has an active role in water formation, which appears to be more complicated than previously thought. Amorphous water ice layers are the matrices where complex organic prebiotic species may be synthesized. This experiment opens up the field of a little explored complex chemistry that could occur on interstellar dust grains, believed to be the site of key processes leading to the molecular diversity and complexity observed in our universe.**

.

**Introduction**

Water, the spring of life (*1*), is the most abundant molecule in biological systems, and it is almost certainly of extraterrestrial origin. Water has been detected, in gaseous or solid form, in numerous astrophysical environments such as planets, comets, interstellar clouds and star forming regions where strong maser emission can be also observed (*2, 3*). Amorphous water ice was directly detected in dark interstellar clouds through infrared absorption (*4*). During the formation of stars deep inside molecular clouds, gas and dust become part of the infalling material feeding the central object. Part of this gas and dust grains, covered with icy mantles (mainly composed by water), ends up in the rotating disks surrounding young stars and forms the basic material from which icy planetesimals and later planets, together with comets in the external regions, are formed (*5*). While the means of delivery of water to Earth remains a subject of debate (*6*) (primordial, cometary…), the synthesis of water in the universe is a fundamental link in establishing our origins.

Water molecule formation in the gas phase is not efficient enough to reproduce the observed abundances in dark clouds, especially in its solid form (*7, 8*); therefore water ice must form directly on the cold interstellar grains and not as a condensate after being formed in the gas phase. A complete review of the processes involved both in the gas and solid phase has been recently published (*9*).

It was suggested many years ago that interstellar dust grains act as catalysts (*10, 11*). Starting from simple atoms or molecules such as H, O, C, N, CO…, grains are believed to be chemical nanofactories on which more complex molecules are synthesized leading eventually to prebiotic species produced concurrently by surface reactions and by UV photons and cosmic rays irradiation, as already shown long ago (*12, 13*). The most refractory species remain on the grain surface, building up a so called "icy mantle", while the most volatile may be released in the gas phase upon formation. Such mantles, having an average thickness of some hundreds of monolayers, are mainly composed of water, the most abundant solid phase species in the Universe. Under dark cloud conditions, except for the very first monolayer that can grow on bare silicate or carbonaceous grains (*14*), most of the water molecules could be subsequently synthesized on a surface mainly composed of water. Thus most of the water molecules should be formed on a water ice substrate in dark interstellar clouds. The experimental results we report here concern precisely with water molecule formation on a water ice substrate at 10K.

Up to date no experimental investigations of water formation under conditions relevant to astrophysics have been performed. Chemical models including water formation have been proposed years ago (*15*). In the case of $H_2O$ formation, two recent models, have explored two different chemical routes. After OH formation on the surface, one model (*16*) proposes the OH+H reaction to occur, whereas in another approach (*17*) this reaction is not considered efficient and the reaction chain $OH+H_2$ or $H_2O_2+H$ is preferred. Both models reproduce reasonably satisfactorily the observed abundances, including the fascinating aspect of deuterium fractionation.

The aim of this first attempt to synthesize water in conditions close to those encountered in dense clouds is to discover if water formation can occur even at temperatures as low as 10K, and possibly which chemical path(s) is(are) actually active.

**Experimental procedures**

Experiments have been performed with the FORMOLISM set-up (*18*). A copper disk that can be temperature controlled in the 8-800K range is maintained under UHV conditions. The ice substrate, on which water ice formation is studied, is prepared in two steps. 100 layers of non porous amorphous $H_2O$ ice are first deposited at 120K (*19*), then an outer layer is made of 10 layers of porous ice deposited at 10K. The first layer isolates the substrate from the copper substrate (*20*). This double ice layer is annealed at 90K prior to any experiments in order to avoid collapse of the pores between 10 and 80K in the subsequent Temperature Programmed Desorption (TPD) experiments (see below). During the experiments, the species $D_2$, O, $O_2$ and $O_3$ therefore evaporate before any rearrangement in the porous structure of the ice template. The ice substrate is still porous (*19*) and mimics an amorphous ice processed by UV and cosmic rays (*21*) that constitutes the icy mantle of interstellar grains.

Using an architecture with 2 separate channels (*22*), two atomic beams of O and D are aimed at the ice substrate held at 10 K. The atoms are produced by dissociation of $O_2$ and $D_2$ in microwave discharges. Dissociation rate of the oxygen beam is typically 40%, meaning that for 100 $O_2$ molecules that initially feed the discharge, 60 $O_2$ molecules and 80 O atoms will finally reach the ice substrate. A few traces of residual gases, namely CO, $CO_2$, OH, and $H_2O$ are present in the beam. They are detected when the movable Quadrupole Mass Spectrometer (QMS) intercepts the beam. Neither $O_3$ nor deuterated compounds could be detected. The D beam has a 60% molecules dissociation rate, and no UV photons from the $D_2$ discharge plasma can reach the ice sample in that case. The O and D beams have respectively a flux of $10^{12}$ atoms/cm²/s and $4\ 10^{12}$ atoms/cm²/s. After simultaneous or alternating injection of both beams, TPD is performed by linearly increasing at 20K/minute the temperature of the sample from 10 K to 200 K and monitoring the desorbing species using the QMS.

The contribution from the water substrate itself could complicate the analysis because it represents more than 99% of the desorbing molecules and since ~0.2% of terrestrial water is composed of HDO. In order to evaluate the contribution of the ice template itself to the measurement of masses other than mass 18, we have monitored with the QMS during the water substrate deposition. This enabled us to subtract the substrate contribution. Finally, we have used both $^{18}O_2$ and $^{16}O_2$ as precursors for the oxygen atomic beam in order to obviate possible mass ambiguities. We never employed high surface concentration of gas and have worked with the monolayer coverage.

**RESULTS**

Several different experiments have been performed and are here reported.

**i) Oxygen beam in absence of D**

The solid line in Fig. 1 shows desorption profile of mass 32 after 5 minutes of O beam injection on the ice sample held at 10K. Desorption occurs in two temperature regions, one between 32 and 55K and the other between 60 and 85 K. The $O_2$ desorption occurs in the first region, as has been verified by using the beam with the discharge off (pure $O_2$). The peak in the second region is proportional to the $O_3$ desorption signal monitored simultaneously at mass 48 (see the inset in Fig. 1) and appears at the same temperature as previously observed (*23*). This peak at mass 32 represents the cracking fraction of $O_3$ in the electron impact ionizer of the QMS and is therefore an indirect measurement of ozone desorption.

O atoms desorption at mass 16 has not been observed. We can conclude that O atoms form $O_2$ and $O_3$ on the water ice substrate in the absence of other reactants. We cannot distinguish if such a formation process occurs at 10 K or if it is thermally induced during the heating ramp.

**ii) Reactivity of D in the absence of Oxygen**

We observe first that injection of D atoms on the water substrate does not produce any deuterated water (HDO) molecules as shown on Fig. 2 (thick line). D forms $D_2$ in absence of oxygen compounds and does not react with the substrate. The isotopic exchange that has been observed in comparable experiments with methanol (*24*) is inefficient with water.

**iii) Reactivity of D in the presence of Oxygen**

On the other hand, the injection of D atoms following O injection has a great effect on the oxygen molecules desorption. The circles in Fig. 1 show the desorption profile of mass 32 after 5 minutes exposure of O beam followed by 5 minutes of D beam. Most of the mass 32 signal has disappeared, implying that either oxygen molecules and ozone had reacted with D if they were already formed or oxygen atoms had reacted with D atoms before forming most of $O_2$ and $O_3$. We conclude that D reacts quite rapidly with $O_2$ and O or $O_3$ at 10 K. The D and oxygen recombination reaction is in competition with $D_2$ formation.

A second experiment was made with $O_2$ beam injection followed by D injection. It leads to the same conclusion. The $O_2$ desorption peak disappears after D injection. This experiment was performed in the total absence of UV photons emitted by the dissociation sources, therefore showing that the reactivity of D and $O_2$ is not photo-induced.

**iv) Oxygen atoms reactivity with $D_2$**

Desorption spectra after O injection followed (or preceded) by that of $D_2$ are similar to desorption spectra obtained after O injection alone. Under our experimental conditions (reaction time ~ 1 minute), $D_2$ is not reactive at 10K. In our experiments, D atoms are a requisite for reaction induction.

**v) Desorbing molecules after O + D deposition (in presence of $O_2$ and $D_2$)**

Circles and thin line in Fig. 2 show the desorption profiles, simultaneously recorded, of respectively mass 19 (HDO) and mass 20 ($D_2O$). The desorption temperature range is the same for the 3 water isotopologues. We observe that much more HDO is produced than $D_2O$. Oxygen, deposited in two separate experiments in the form of O or $O_2$, desorbs mostly in the form of HDO. Some $O_2$ and $O_3$ remains. A few % desorb as $O_2D$ (or $H_2O_2$) and ~1% as $D_2O$.

**vi) Beams with $^{18}O$ + D**

We have also performed experiments using $^{18}O$. We observe that the shape of the mass 19 desorption curve is not similar to that of water at mass 18. It is thus reasonable to envisage two components, $HD^{16}O$ and $^{18}OH$. We therefore conclude that an important part of $^{18}O$ is locked in $^{18}OH$. We observe that $^{18}O$ also desorbs in the form of $^{18}OD$ (mass 20). Fig. 3 shows the desorption profile of mass 21 and 22, performed after 10 minutes of simultaneous D and $^{18}O$ injection. Mass 21 can only be assigned as $HD^{18}O$ and mass 22 as $D_2^{18}O$. We can see that even if these peaks are not as intense as expected, new water molecules, labelled by $^{18}O$, are formed, and that again the deuterated isotopologue HD$^{18}$O dominates. This is the proof that water formation takes place on the surface.

**Discussion**

HDO is the major product of reaction when either $^{16}$O or $^{18}$O injections are performed together with D on the substrate. D comes from the D atomic beam but H atoms are obviously provided by the water substrate. This implies that the water substrate itself plays a major role in the reactions occurring on the surface. This fact has never been envisioned in the modelling of water synthesis in the astrophysical contexts. The desorption of water containing $^{18}$O in the case of isotopically labelled experiments (Fig 3) shows that oxygen is not only a catalyst in a kind of indirect water deuteration reaction (*24*), but is a real partner. We suggest that the following mechanism could occur on the surface. We think that a possible explanation of our results goes through the formation of $H_2O_2$ as intermediary which reacts with D atoms to produce HDO+OH. Then OH and OD (coming from O+D or $O_2$+D) accumulate and can react together on the surface to form $HDO_2$ or $D_2O_2$. Both react efficiently with D. Only a little $D_2O$ is finally formed under our low coverage conditions. We can also conclude that the OD+D reaction is inefficient at 10 K. If this reaction does take place, much more $D_2O$ should be observed, and only a little OH or OD should desorb.

In dense clouds, hydrogen is almost exclusively in its molecular form. H atoms represent a small fraction of hydrogen and are not dependent on the density in the clouds (*16*). By contrast, O is mostly in its atomic form. Even if the cosmic elementary abundances give a ratio of O/H ~ 3 10$^{-4}$ (*25*), it is believed that, in dense clouds, O/H can vary between 0.1 and 100. Since O forms $O_2$ and $O_3$ in clouds where O/H>1, all forms should be present, at least transiently. Therefore our experimental conditions, where O is mixed with $O_2$, are relevant. The experiment performed with $O_2$+D is very instructive since reactions take place in the absence of photons. It demonstrates that water formation can be effective in photo-shielded environments such as dense clouds (even if secondary UV photons induced by cosmic rays are present). Experimentally we observe that $O_2$ is easily destroyed by D atoms. It may explain why no positive detection of $O_2$ has been reported (*5*). By contrast we observe some OH formation. We believe that in astrophysical scenarios, OH is not present as observed in our experiments for 3 main reasons. Firstly, if OH accumulates, above a certain surface concentration it could self react and subsequently form water; this could be experimentally tested by using longer O and D injection durations. Secondly it is also possible that OH reacts with $H_2$ and forms $H_2O$, thus releasing one H available for hydrogenation of the adsorbed species as suggested by models (*15*). We cannot easily check experimentally the presence of the $OH+H_2$ reaction since if it occurs, it should be on a very long time scale, as suggested by the activation barrier estimated to be 2600 K (*7*) . Finally, other molecules, such as CO, can react with the OH radicals. This opens the door to a new and almost unexplored chemistry at 10 K. Alcohol, formic acid and other compounds are observed in icy mantles. Their release into the gas phase is inferred in hot cores during an early stage of the star formation process. Therefore a more complete understanding of the increasing complexity of molecules formed on interstellar dust grains is a key to constrain the physical conditions leading to star and planet formation.

**Conclusion**

The first synthesis of water has been performed under conditions relevant to dark cloud environments. HDO is produced with high efficiency by D and O beam injection on a water ice sample held at 10K,. The water sample which mimics the icy mantle of interstellar dust grains, has an active role in water formation. Consequently the formation mechanism appears to be different than previously thought. This experiment opens up a new and unknown field of complex catalysis at low temperatures. Interstellar ices are not only composed of water, and other species can act as catalyst or reactant, which calls for future experiments. By adding other compounds (CO, $NH_3$…) the path to chemical complexity is now open, giving astrochemistry a stronger link to astrobiology.(26)

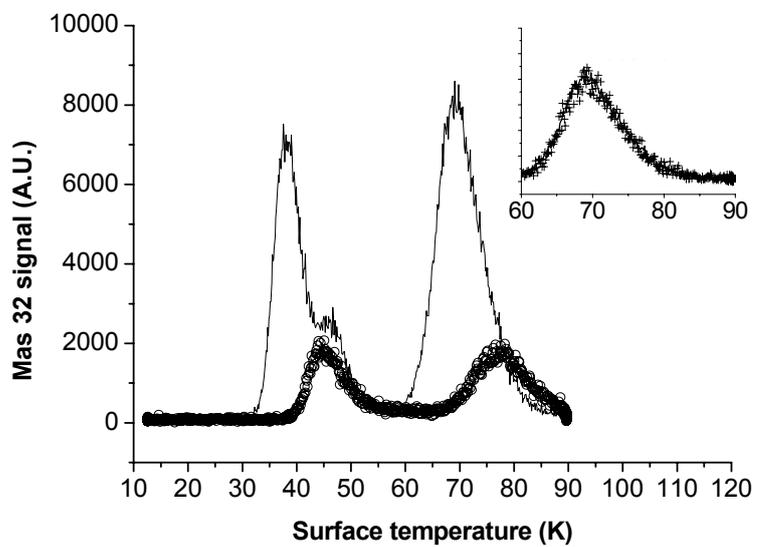

Fig. 1: Desorption profiles of mass 32, after 5 minutes of O injection, followed (ooo) or not followed (——) by 5 minutes of D exposition. Ice sample is held at 10 K during expositions. Insert: comparison of rescaled Mass 32 (——) and Mass 48 (+++) ($O_3$) signals after 5 minutes of O injection only.

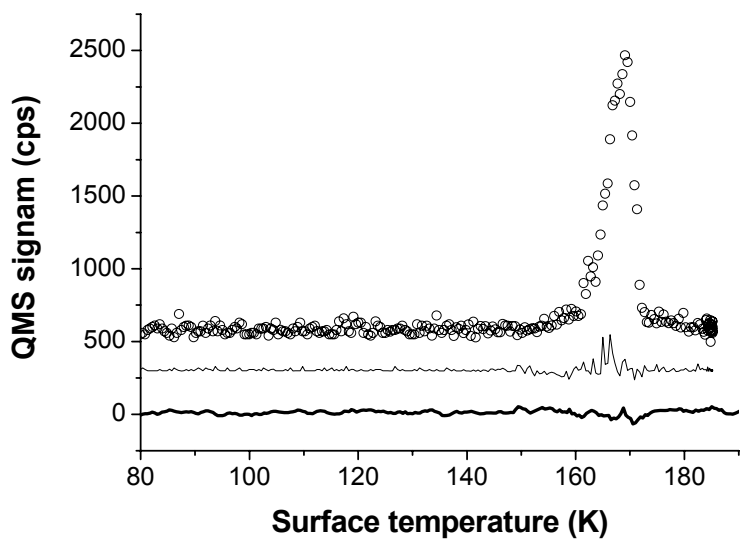

Fig. 2 – Desorption profile of mass 19 (—) after 10 minutes exposition of D beam. Simultaneous records of desorption profile of mass 19 (ooo) and mass 20 (—) after 5 minutes of simultaneous injection with $^{16}$O and D atoms. The 3 curves are vertically shifted for clarity.

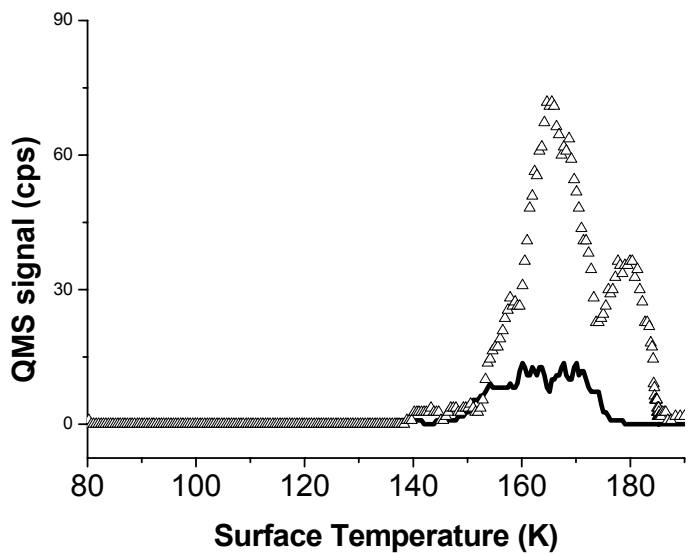

Fig. 3 : HD$^{18}$O (mass 21 ΔΔΔ) and D$_2$$^{18}$O (mass 22 —) desorption profiles after 10 minutes of simultaneous injection with $^{18}$O and D beams.

1. A. Brack, in *Astrobiology: The quest for the conditions of life* G. Horneck, C. Baumstark-Khan, Eds. (Physics and astronomy online library, Springer, Berlin, 2002) pp. 79.
2. P. Ehrenfreund *et al.*, *Planet. Space Sci.* **51**, 473 (2003).
3. E. Dartois, *Space Science Reviews* **119**, 293 (2005).
4. A. Leger *et al.*, *A&A* **79**, 256 (1979).
5. E. F. van Dishoeck, *Annu. Rev. Astron. Astrophys.* **42**, 119 (2004).
6. A. Morbidelli *et al.*, *Meteoritics & Planetary Science* **35**, 1309 (2000).
7. B. Parise, C. Ceccarelli, S. Maret, *A&A* **441**, 171 (2005).
8. C. Ceccarelli, P. Caselli, E. Herbst, A. G. G. M. Tielens, E. Caux, in *Protostars and Planets V* B. Reipurth, D. Jewitt, K. Keil, Eds. (U. Arizona Press, 2007) pp. 47.
9. A. G. G. M. Tielens, *The Physics and Chemistry of the Interstellar Medium* (Cambridge University Press, 2005), pp. 1-495.
10. J. H. Oort, H. C. van de Hulst, *BAN* **10**, 1870 (1946).
11. H. C. van de Hulst, *Recherches Astronomiques de l'Observatoire d'Utrecht* **11**, 2 (1946).
12. W. Hagen, L. J. Allamandola, J. M. Greenberg, *Astrophysics and Space Science* **65**, 215 (1979).
13. V. Pirronello, W. L. Brown, L. J. Lanzerotti, K. J. Marcantonio, E. H. Simmons, *ApJ* **262**, 636 (1982).
14. R. Papoular, *Mon. Not. R. Astron. Soc.* **362**, 489 (2005).
15. A. G. G. M. Tielens, W. Hagen, *A&A* **114**, 245 (1982).
16. P. Caselli, T. Stantcheva, O. M. Shalabiea, V. I. Shematovich, E. Herbst, *Planetary and Space Science* **50**, 1257 (2002).
17. B. Parise, PhD Thesis, CESR (Toulouse, France) (2004).
18. L. Amiaud *et al.*, *J. Chem. Phys.* **124**, 094702 (2006).
19. G. A. Kimmel, K. P. Stevenson, Z. Dohnalek, R. S. Smith, B. D. Kay, *J. Chem. Phys.* **114**, 5284 (2001).
20. I. Engquist, I. Lundström, B. Lieberg, *J. Phys. Chem.* **99**, 12257 (1995).
21. M. E. Palumbo, *A&A* **453**, 903 (2006).
22. V. Pirronello, C. Liu, L. Shen, G. Vidali, *ApJ* **475**, L69 (1997).
23. F. Borget, T. Chiavassa, A. Allouche, J. P. Aycard, *J. Phys. Chem. B* **105**, 449 (2001).
24. A. Nagaoka, N. Watanabe, A. Kouchi, *ApJ* **624**, L29 (2005).
25. D. M. Meyer, M. Jura, J. A. Cardelli, *ApJ* **493**, 222 (1998).

26. Acknowledgements: We thank E. Herbst, J. Cernicharo and S. Leach as well as D. Field and L. E. Kristensen for useful discussions and comments. We acknowledge support of the national PCMI program funded by the CNRS, as well as strong financial support from the Conseil Régional d'Ile de France (SESAME program) and the Conseil Général du Val d'Oise.

Dear Editor,

once the surprise has been overcome, we would like to make a short reply and a few comments about the two reviews.

As far as referee 2 is concerned, we are very grateful for his acknowledgement of the reality and relevance of our discovery and for the very deep reading of our paper as well as the serious work he has done trying to improve the paper through his comments. Nevertheless it clearly seems to us that he was not aware that we had sent a report, due to the novelty of our discovery, and not a full paper which has to appear later. We therefore could not address all the requests that the referee made in the space available for the limited space a report allows. The fact that it was a report was however clearly stated in our cover letter.

The first referee on the contrary tried only to kill our paper for whatever reason (may be a competitor and you will have the chance to see it in the future); alternatively he didn't have a good understanding of the subject.

We did not cite Hiraoka et al 1998 for 2 main reasons:
1) $N_2O$ is the ice matrix in which water formation is claimed to have been obtained from H reacting with O kept in this matrix. But this matrix as well as O were obtained from a $N_2O$ activated plasma (DC discharge 2-10 kV). Therefore the ice mixture obviously contains, in addition to O (probably quite energetic), also very reactive species such as N and NO.
2) such a $N_2O$ ice (containing in addition to O, N etc…) is not relevant at all to astrophysics and astrochemistry on interstellar dust grains (See for instance Tielens, Cambridge Edt., 2005 and more in general the literature about icy mantles on grains).
An experiment of this type needs to be a simulation as close as possible of what occurs in space, not just an attempt to form molecules in any way. Our experiment is showing that it is possible to form a water ice mantle on interstellar grains (monolayer after monolayer) by reactions with accommodated atomic hydrogen and atomic oxygen, the paper of Hiraoka et al does not.
The remark that to publish another paper one should wait until in-situ detection of water formation performed at 10K seems not to consider the fact that water in our experiment must have formed at temperatures not higher than approximately 15 - 20 K, because H atoms do not stay on water ice surface for long at temperatures higher than those values. Thus, the combination of TPD and mass spectrometric detection of the products is a clear evidence that water is formed in the solid phase in conditions very similar to those encountered in dense interstellar clouds and can give rise to the growth of the observed icy mantles on grains.
It is also quite strange that the first referee seems not to be aware of such elementary facts.
Finally the 1998 paper by Hiraoka et al was announcing as a confirmation a paper on hydrogenation reactions of solid $O_2$ and $O_3$ which does not seem to have been performed and published since.

As a final remark we should like to let notice to the Editor that such opposite judgments between the two referees might deserve a different action.

In the light of these comments we would like to ask you to change the first referee and to reconsider our report, or alternatively, an even shorter letter presenting only the simple evidence of our experimental results.

Best regards

The authors